\documentclass[reprint,aps,prb,superscriptaddress,a4paper,showpacs,showkeys,floatfix]{revtex4-1}

\setcitestyle{numbers,square}

\usepackage{amssymb,amsmath,bm,dcolumn,braket} 
\usepackage{rotating}
\usepackage{url}
\usepackage{color}
\usepackage[normalem]{ulem} 

\usepackage{graphicx,epsfig,epstopdf}
\usepackage[caption=false]{subfig}

\newcommand{\ms}{\mbox{$\mu_{\mathrm{spin}}$}}
\newcommand{\mo}{\mbox{$\mu_{\mathrm{orb}}$}}
\newcommand{\mB}{\mbox{$\mu_{B}$}}
\newcommand{\nfe}{\mbox{$N_{\text{Fe}}$}}

\newcommand{\kkr}{{\sc sprkkr}} 
\newcommand{\wien}{{\sc wien}2k}

\newcommand{\ea }{{\it et al.}}

\begin{document}


\title{Local environment effects in the magnetic properties and electronic
  structure of disordered FePt}

\author{Saleem Ayaz \surname{Khan}} \affiliation{New Technologies
  Research Centre, University of West Bohemia, Univerzitn\'{\i}~2732,
  CZ-306~14~Pilsen, Czech Republic}

\author{Jan \surname{Min\'{a}r}} \affiliation{New Technologies
  Research Centre, University of West Bohemia, Univerzitn\'{\i}~2732,
  CZ-306~14~Pilsen, Czech Republic} \affiliation{Department Chemie,
  Universit\"{a}t M\"{u}nchen, Butenandtstr.~5-13,
  D-81377~M\"{u}nchen, Germany}

\author{Hubert \surname{Ebert}} \affiliation{Department Chemie,
  Universit\"{a}t M\"{u}nchen, Butenandtstr.~5-13,
  D-81377~M\"{u}nchen, Germany}

\author{Peter \surname{Blaha}} \affiliation{Institute of Materials
  Chemistry, TU~Vienna, Getreidemarkt~9, A-1060~Vienna, Austria}

\author{Ond\v{r}ej \surname{\v{S}ipr}} \affiliation{New Technologies
  Research Centre, University of West Bohemia, Univerzitn\'{\i}~2732,
  CZ-306~14~Pilsen, Czech Republic} \affiliation{Institute of Physics
  ASCR v.~v.~i., Cukrovarnick\'{a}~10, CZ-162~53~Prague, Czech
  Republic }


\date{\today}

\begin{abstract}
Local aspects of magnetism of disordered FePt are investigated by ab
initio fully relativistic full potential calculations, employing the
supercell approach and the coherent potential approximation (CPA). The
focus is on trends of the spin and orbital magnetic moments with
chemical composition and with bond lengths around the Fe and Pt
atoms. A small but distinct difference between average magnetic
moments obtained when using the supercells and when relying on the
CPA is identified and linked to the neglect of the Madelung potential
in the CPA.
\end{abstract}


\maketitle


\section{Introduction}   \label{sec-intro}

When dealing with substitutional alloys, it is necessary to use
approximations to simulate the random occupation of sites. There are
several ways to achieve this.  One can use supercells to include as
many different local configurations as possible.  Or one can use a
mean field approach to simulate the disorder by a suitably chosen
auxiliary effective medium.  Both of these limiting approaches can be
combined, as in the locally self-consistent Green function formalism.
Each method has its advantages and limitations.

The mean field approach such as the single site coherent potential
approximation (CPA) is computationally convenient and the random
disorder itself is treated very efficiently. However, single-site
methods neglect the influence of fluctuations in the local environment
\cite{GSB+84,AJ+98}. Another problem is that the Madelung contribution
to the alloy potential cannot be included within the standard CPA
\cite{JP+93,RSK+02}.  A lot of effort was devoted to tackle this issue
and important advances in this respect have been made
\cite{Johnson+93, Magri+90, Korzhavyi+95, Saha+96, Ujfalussy+00,
  Bruno+02, Abrikosov+98, Ruban+02}.

Theoretical investigations on the electronic structure of alloys
mostly concerned topics like ordering, phase stability
\cite{SBD+01,Sluiter+06} and systematic dependence of various
properties on the alloy composition \cite{ADK+88,HD+98,KDT+08,YPA+15}.
Studies of local environment effects focused on charge transfer and
energetics \cite{RAS+95,AJ+98,BZM+02,KMS+06,KSS+16}.  Few studies
dealt also with short-range order effects on magnetic moments
\cite{Paudyal+04,PRJ+12}.  Comparison between theory and experiment
shows, nevertheless, that the CPA is often able to describe the trends
of physical properties with alloy composition very well
\cite{ADK+88,KSO+06,KDT+08,LKE+09,YPA+15}. Here, we want to further
investigate magnetic properties and specifically we want to focus on
the relation between the local magnetic moments and local atomic
environments.  To get a comprehensive view, we use the supercell as
well as the CPA technique.  Our calculations employ the full potential
and include relativistic effects such as spin-orbit coupling (SOC).

We focus on the Fe$_{0.5}$Pt$_{0.5}$ substitutional alloy.  This
system attracted considerable attention in the past.  Perlov
\ea\ \cite{Perlov+98} performed a systematic study to investigate
electronic structure and magneto-optical properties of disordered FePt
alloys.  Kharoubi \ea\ \cite{Kharoubi+13} investigated the electronic
structure, the complex Kerr angle and the magnetic moments for ordered
and disordered FePt multilayers and performed a complete analysis of
the strong Kerr rotation with respect to photon energy.  Paudyal
\ea\ \cite{Paudyal+04} calculated the electronic structure and
magnetic properties of ordered and disordered FePt, CoPt and NiPt
alloys. Their main concern was a comparison between the ordered and
disordered phases and the variation of the magnetic moments with alloy
composition.  Sun \ea\ \cite{Sun+06} explored magnetic moments and
magnetic circular dichroism (MCD) of ordered and disordered
Fe$_{0.5}$Pt$_{0.5}$ films using a fully relativistic KKR code. They
confirmed that the spin magnetic moment of Fe is similar for ordered
and disordered films. However, the orbital magnetic moment measured
with MCD is found to be larger than predicted by theory \cite{Sun+06}.

The aim of this study is to complement earlier research on FePt and
other alloys by investigating local variations of the electronic and
magnetic structure.  Specifically, we want to focus on the effect of
the chemical composition of the first coordination shell and on the
effects of structural relaxations.  As we are interested in local
effects, we will use mostly the supercell approach.

An important part of our study will be a comparison between CPA and
supercell-based calculations.  On the one hand, one can view this as a
check whether the choice of the supercells is representative enough
for simulating substitutional disorder.  On the other hand, one can
view this as a check whether local environment fluctuations lead to
a significant effect on aggregate properties such as average magnetic
moments.

The paper is arranged as follows. We start by explaining the
computational details. Then we present our results regarding the
dependence of the local magnetic moments on the chemical composition
of nearest neighbors.  The difference between supercells and CPA
results is interpreted by analyzing the effect of the Madelung
potential on magnetic moments.  We also discuss changes of local
magnetic moments caused by variations of the bonding lengths resulting
from structure relaxation. We discuss our results and summarize the
conclusions in the final sections.


\section{Methodological framework}

\subsection{Crystal structures simulating the disorder}

\label{sec-methodology}

We want to study local environment and structure relaxation effects in
disordered FePt, therefore we employ the supercell technique.
However, such an approach can be computationally expensive and
cumbersome, because a very large supercell may be needed to represent
a disordered systems with sufficient accuracy.  Hence one tries to
make the supercells as small as possible while making sure that the
distribution of atoms among the sites represents the disorder
sufficiently.

The concept of a special quasirandom structure (SQS) developed by
Zunger \ea~\cite{Zunger+90} is motivated by efforts to generate the
supercells in an efficient way.  The basic idea is to create periodic
structures so that they have the same pair and multi-site correlation
functions as random alloys up to a certain coordination shell. The
distribution of local environments created along these lines is
representative for a random alloy, at least up to the first few
coordination shells.  In addition a weighted average should be taken
over various SQS's to smear out artificial effects of the periodicity
of the supercell.

\begin{figure}
\subfloat[\label{SQS-4}]{%
\includegraphics [width=0.3\linewidth]{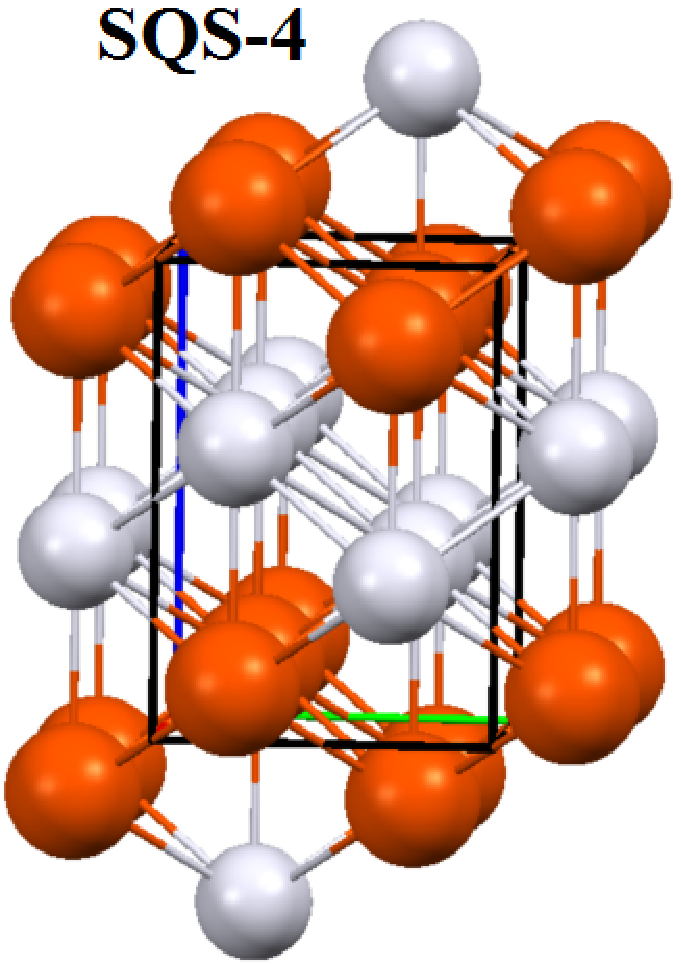}} \hfill
\subfloat[\label{SQS-8}]{%
\includegraphics [width=0.54\linewidth]{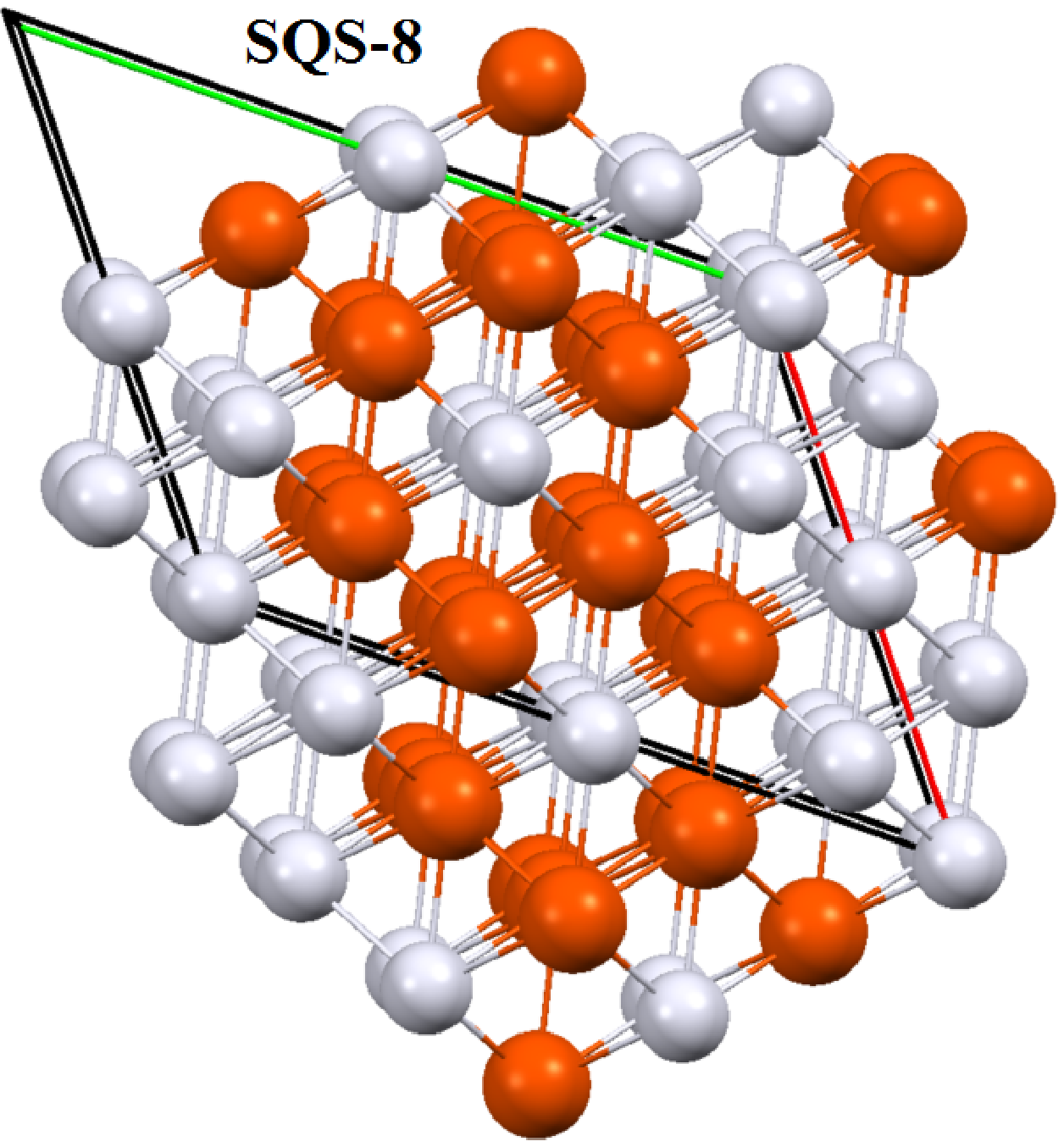}}\hfill
\subfloat[\label{SQS-16}]{%
\includegraphics [width=0.4\linewidth]{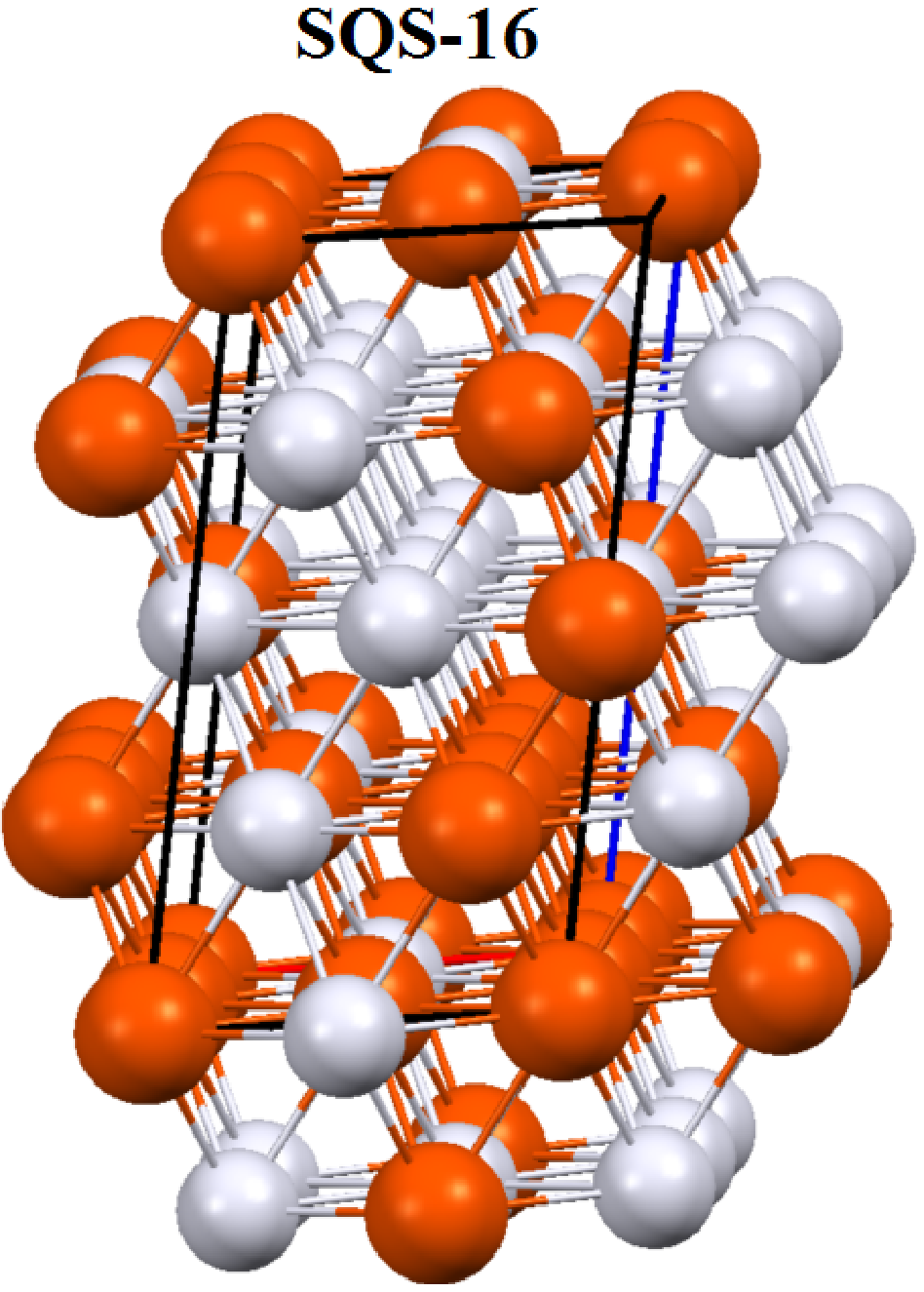}}\hfill
\subfloat[\label{SQS-32}]{%
\includegraphics [width=0.6\linewidth]{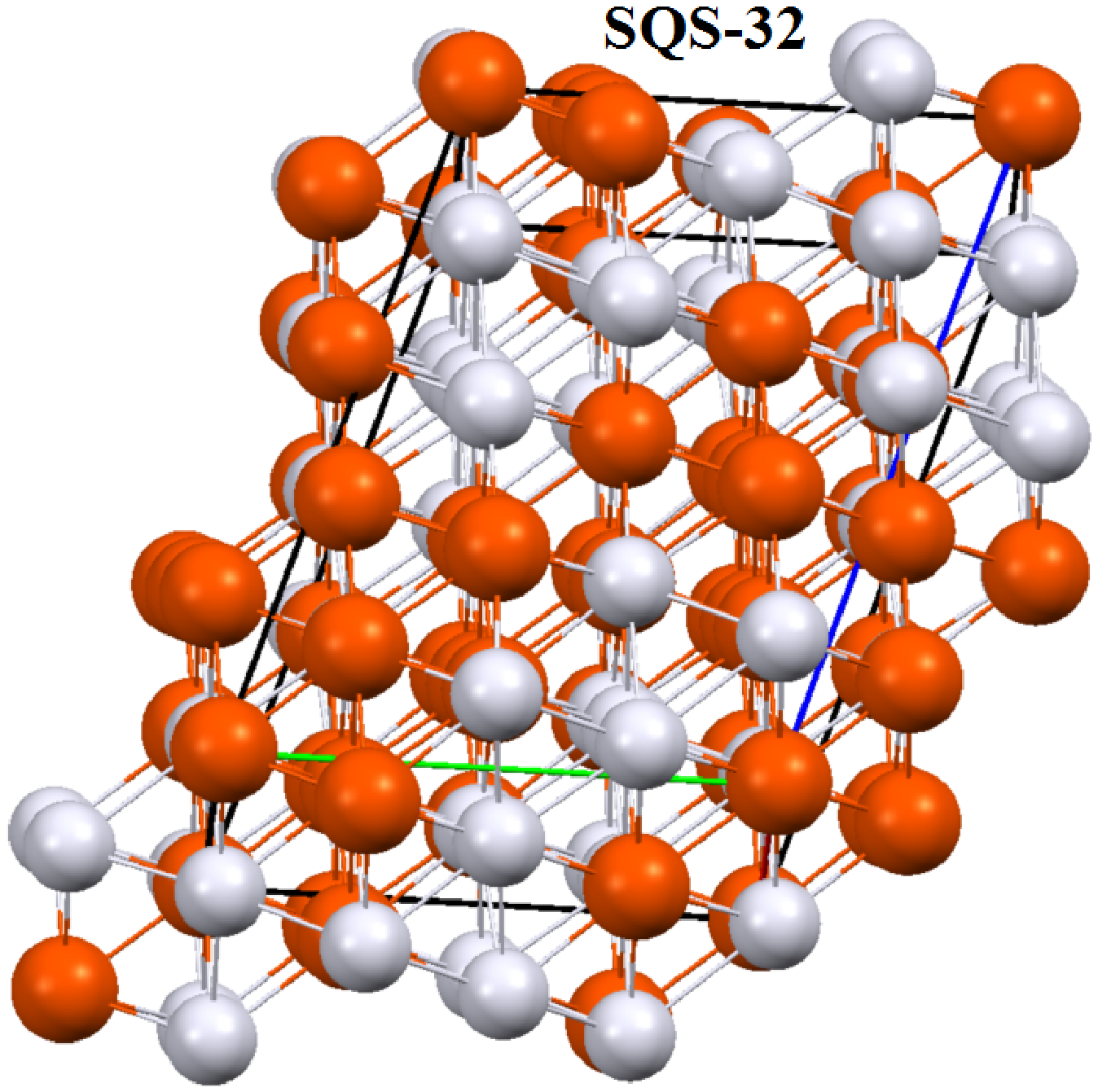}}
\caption{Structure diagrams of the SQS's used in this
  study. \label{unit-cell}}
\end{figure}

The crystal structure of FePt disordered substitutional alloy (50:50
concentration) is A1 (fcc) with a lattice constant
$a=3.807$~\AA\ \cite{Lub+07}. We model it by SQS's with an underlying
fcc lattice, containing $N=4$, 8, 16, and 32 atoms per unit cell.  The
crystallographic data for SQS-4 are taken from Su
\ea\ \cite{Su+10}, the data for SQS-8, SQS-16, and SQS-32 are
taken from Shang \ea\ \cite{Shang+11}.  Note that the SQS-8 structure
we use is equivalent to the SQS-8$b$ structure of Lu
\ea\ \cite{Lu+91}.  Structural diagrams for our SQS-$N$ systems are
shown in figure~\ref{unit-cell}.

For comparison, we performed calculations for ordered FePt as well.
We considered either a model L1$_{0}$ structure isostructural with the
A1 structure ($c/a=1$, this could be interpreted also as the SQS-2
structure \cite{Wei+90}), or the L1$_{0}$ structure corresponding to
true ordered FePt system ($a$=3.849~\AA, $c$=3.714~\AA, $c/a=0.965$).

\subsection{Computational details}

We used two computational methods, namely, the FLAPW method as
implemented in the \wien\ code \cite{Blaha+01} and the fully
relativistic full-potential multiple scattering KKR (Korringa
Kohn-Rostoker) Green function method \cite{sprkkr+11} as implemented
in the \kkr\ code \cite{sprkkr+12}.  We treated the Fe $3p$, $3d$,
$4s$ and Pt $5p$, $5d$, $6s$ states as valence states and,
accordingly, the Fe $1s$, $2s$, $2p$, $3s$ and Pt $1s$, $2s$, $2p$,
$3s$, $3p$, $3d$, $4s$, $4p$, $4d$, $4f$, $5s$ states as core states.

For the \wien\ calculations, the crystal is divided into non
overlapping muffin-tin (MT) spheres and the interstitial region
between them.  The wave function inside a MT sphere is expanded in
terms of atomic-like wavefunctions, with the expansion controlled by
the angular-momentum cutoff $\ell_{\text{max}}^{\text{(APW)}}$.  The
wave function in the interstitial region is expanded in terms of plane
waves, with the plane wave cutoff determined by the product
$R_{\text{MT}}K$, where $R_{\text{MT}}$ represents the muffin-tin
sphere radius and $K$ the magnitude of the largest wavevector. We use
$\ell_{max}^{\text{(APW)}}$=10, $R_{\text{MT}}$K$_{max}$=8.0,
R$_{\text{MT}}$(Fe)=2.2~a.u.\ and
$R_{\text{MT}}$(Pt)=2.3~a.u.\ throughout this work.

In the full-potential KKR-Green function calculations, one employs a
multipole expansion of the Green function for which we used a cutoff
$\ell_{\text{max}}^{\text{(KKR)}}$=3. Note that the cutoffs
$\ell_{\text{max}}^{\text{(APW)}}$ and
$\ell_{\text{max}}^{\text{(KKR)}}$ have different meanings within
FLAPW and KKR methods, so their values cannot be directly compared.

Once the Green function components (in KKR calculations) or the wave
functions (in FLAPW calculations) have been determined, the charge
density can be obtained via the $\bm{k}$-space integration over the
Brillouin zone (BZ).  The results presented in this study were
obtained using an integration mesh of 2000 $\bm{k}$-points for the CPA
calculations and for the ordered 
L$1_{0}$ structures, 1000 $\bm{k}$-points for SQS-4, 500
$\bm{k}$-points for SQS-8, 250 $\bm{k}$-points for SQS-16 and 125 $\bm{k}$-points for SQS-32.   All
numbers relate to the full BZ.

For systems of 3$d$ and 5$d$ metals there is a problem whether the
exchange and correlation effects should be dealt with within the local
density approximation (LDA) or within the generalized gradient
approximation (GGA) \cite{Ruban+03,Shang+11}. Our focus is on magnetic
properties, so we used the Vosko, Wilk, and Nusair parametrization
within the LDA scheme \cite{VWN+80}, because benefits of the GGA are
questionable for magnetic studies \cite{BEA96,GOA+00,HSL+02}.
However, when doing structural relaxation we employed the GGA scheme
\cite{PBE96} as well, to enable a comparison. It turns out in the end
that the choice between the LDA and the GGA is not crucial for our
purpose.

When using the \kkr\ code, local magnetic moments (as well as charges)
were evaluated within the Voronoi polyhedra around the atomic sites.
When using the \wien\ code, the local magnetic moments were evaluated
within the muffin-tin spheres.  This difference does not affect our
conclusions because each code is used to investigate different
aspects. We checked that if the MT spheres used in the \wien\ have
maximum radii (so that they touch), very similar values for the
magnetic moments are obtained for the \wien\ and \kkr\ (differences in
the spin magnetic moments \ms\ calculated via both codes are less than
1~\%, differences in the orbital magnetic moments \mo\ are less than
20~\%).  Such a setting of the MT radii would, however, not be
suitable for structure relaxations for which the \wien\ is mainly used
in this work (Sec.~\ref{sec-lengths}).

\section{Results}
\label{Results}


\subsection{Density of states}

\label{sec-dos}

\begin{figure}
\includegraphics[width=0.9\linewidth]{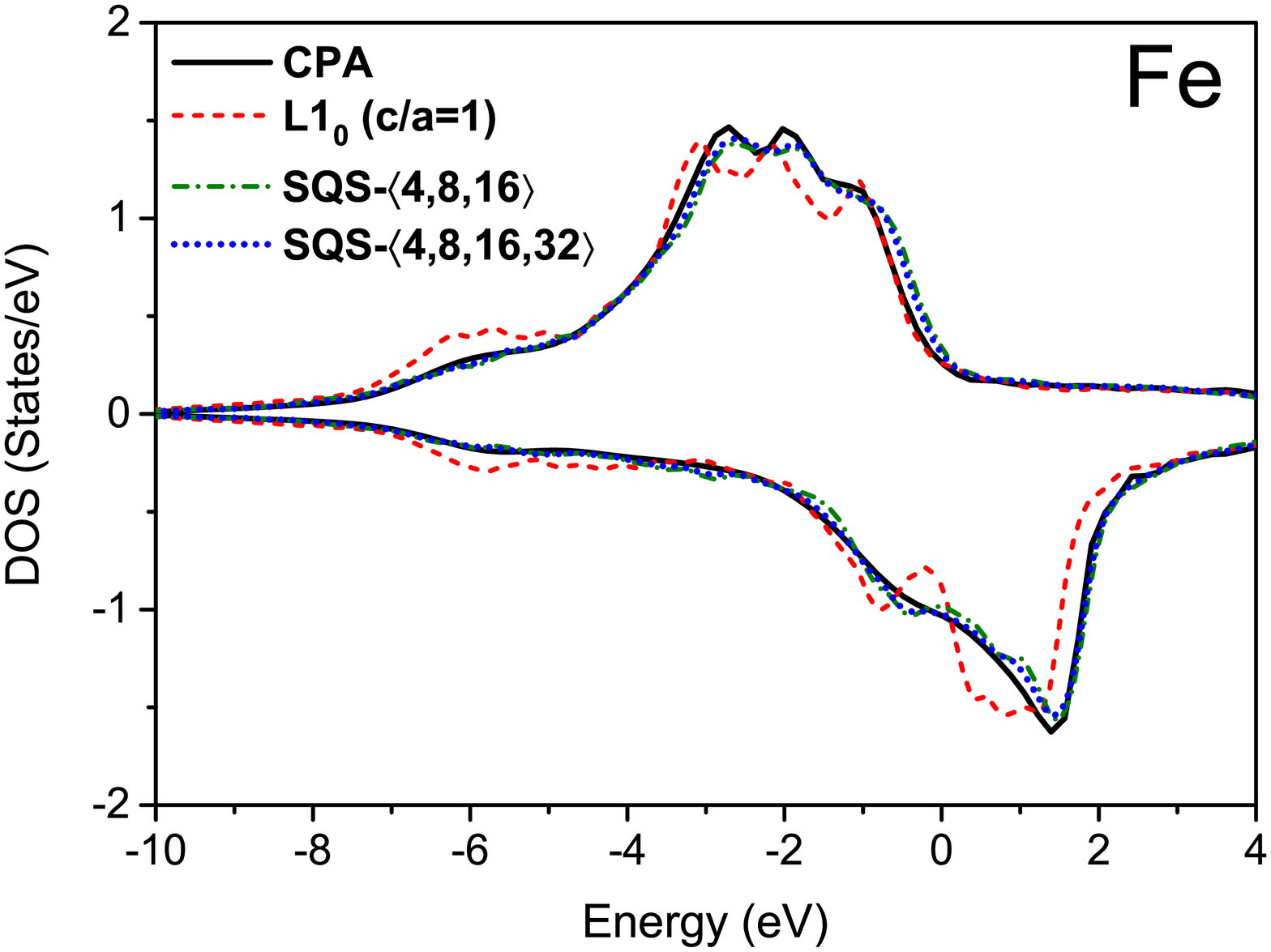}
\includegraphics[width=0.9\linewidth]{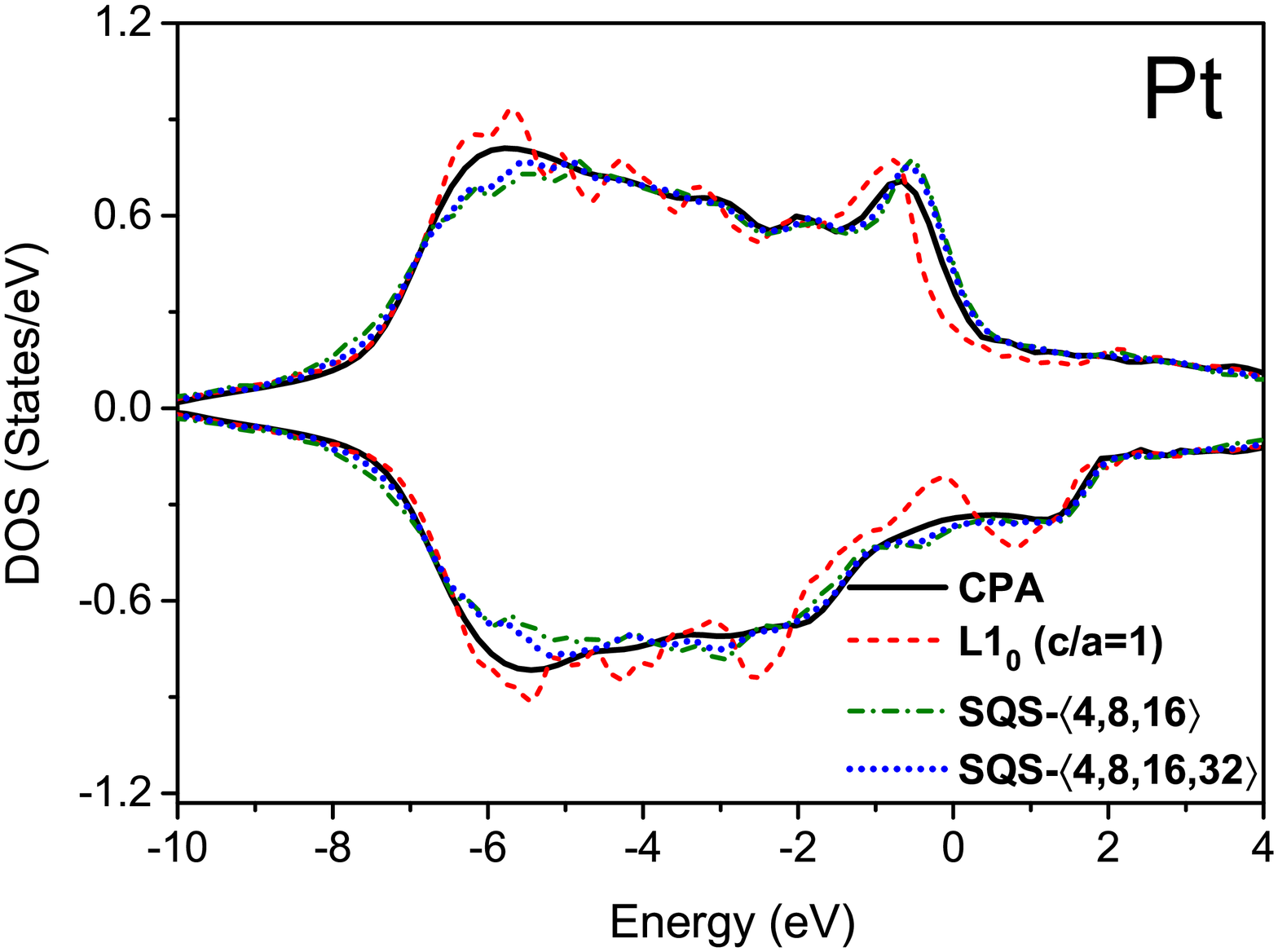}
\caption{Spin-polarized density of states for Fe and Pt sites.  Data
  for ordered L$1_{0}$ structure are compared to data obtained by
  averaging over all sites in the SQS-4, SQS-8 and SQS-16 supercells,
  to data obtained by averaging over all sites in the SQS-4, SQS-8,
  SQS-16 and SQS-32 supercells and to the CPA result. The calculations
  were performed by the \kkr\ code.}
\label{TDOS}
\end{figure}

We begin by investigating the density of states (DOS) and compare the
DOS for ordered structures and supercells with the DOS for disordered
FePt obtained via the CPA.  In particular, we want to monitor how the
DOS averaged over all sites of a given chemical type develops if more
and more sites are included.  The results shown here were obtained via
the \kkr\ code but data obtained for the ordered structures and
supercells via the \wien\ code would look practically the same.

A sequence of DOS curves is presented in Fig.~\ref{TDOS}.  We start
with the DOS for the ordered L$1_{0}$ structure (with c/a=1.0), then
comes the DOS averaged over all sites in the SQS-4, SQS-8, and SQS-16
supercells ($\left<\text{SQS-4,8,16}\right>$), DOS averaged over all
sites in the SQS-4, SQS-8, SQS-16, and SQS-32 supercells
($\left<\text{SQS-4,8,16,32}\right>$) and all this is compared to the
DOS for disordered FePt obtained via the CPA.

One can see that the DOS for ordered L$1_{0}$ FePt and disordered FePt
is different (albeit similar).  When the disorder is simulated by
supercells, the CPA limit is approached quite rapidly; the supercell
and CPA results become practically equivalent for
$\left<\text{SQS-4,8,16,32}\right>$.  This indicates that by averaging
over the SQS's of 4, 8, 16, and 32 atoms, a very good description of
the disorder is achieved.  The DOS for the Fe atoms seems to approach
the CPA data more quickly than the DOS for the Pt atoms but the
difference is not big.

A visual inspection of the DOS curves brings an intuitive insight but
cannot substitute for a quantitative analysis.  In the following
sections, attention will be paid to a careful comparison of integral
quantities such as charges and magnetic moments.



\subsection{Average magnetic moments: comparison between supercells
  and CPA}

\label{sec-aver}

\begin{table}
\caption{Average spin and orbital magnetic moments (in $\mu_{B}$ per
  formula unit) for Fe and Pt atoms in two ordered FePt systems and in
  four SQS's simulating disordered FePt alloy.  Average values over
  all sites in all SQS's are also shown.  The CPA results are
  presented at the bottom.  The data were obtained using the
  \kkr\ code. \label{average-moments}}
\begin{ruledtabular}
\begin{tabular}{lccc}
 & type & 
   $\left<\mu_{\text{spin}}\right>$ & $\left<\mu_{\text{orb}}\right>$ \\ 
\hline\hline 
$L1_{0}$ ($c/a=0.965$)  &  Fe  &  2.829  &  0.065  \\ 
                                       &  Pt  &  0.337  &  0.044  \\ [1.2ex]
$L1_{0}$ ($c/a=1$)         &  Fe  &  2.794  &  0.072  \\ 
                                       &  Pt  &  0.322  &  0.048  \\[1.2ex]
SQS-4                              &  Fe  &  2.843  &  0.043  \\ 
                                       &  Pt  &  0.253  &  0.027  \\[1.2ex]
SQS-8                              &  Fe  &  2.821  &  0.069  \\ 
                                       &  Pt  &  0.286  &  0.051  \\[1.2ex]
SQS-16                            &  Fe  &  2.823  &  0.066  \\ 
                                       &  Pt  &  0.263  &  0.042  \\[1.2ex]
SQS-32                            &  Fe  &  2.816  &  0.073  \\ 
                                       &  Pt  &  0.264  &  0.043  \\[1.2ex]
$\left<\text{SQS-4,8,16,32}\right>$ 
                                       &  Fe  &  2.821  &  0.069  \\
                                       &  Pt  &  0.266  &  0.043  \\[1.2ex]
CPA                                 &  Fe  &  2.903  &  0.070  \\ 
                                       &  Pt  &  0.239  &  0.039  \\
\end{tabular}
\end{ruledtabular}
\end{table}

Table~\ref{average-moments} shows the average spin and orbital
magnetic moments $\left<\mu_{\text{spin}}\right>$ and
$\left<\mu_{\text{orb}}\right>$ for Fe and Pt atoms obtained by
assuming SQS supercell geometries and by employing the CPA.  No
structural relaxation was performed at this stage.  The magnetization
is oriented along the [001] direction of the parental fcc lattice. The
calculations were performed using the \kkr\ code, so a direct
comparison between the SQS and the CPA results can be made.  The data
shown in Table~\ref{average-moments} were obtained for a full
potential but we checked that using it was actually not necessary:
when the atomic sphere approximation (ASA) was applied instead, the
spin magnetic moments increased typically by 1~\% and the orbital
magnetic moments by 2-10~\%.

As found earlier \cite{Paudyal+04,Sun+06}, the difference between
moments for ordered and disordered FePt is evident but not large.  The
variation in $\left<\mu_{\text{spin}}\right>$ between different SQS's
is quite small when going stepwise from $N=4$ to $N=32$.  This is
especially true for the Fe atoms. For the Pt atoms, the relative
deviations are a bit larger but still small.  On the other hand, the
variation in $\left<\mu_{\text{orb}}\right>$ is relatively large for
the same sequence of SQS's.  Again, this variation is larger for Pt
atoms than for Fe atoms.

Remarkably, even for the largest SQS we explore, there remains a small
but distinct difference for the magnetic moments between the supercell
and the CPA approaches. The same applies to the average taken over all
SQS's.  This difference will be subject of a further analysis in
Sec.~\ref{sec-madel}.


\subsection{Dependence of local magnetic moments on the chemical
  composition of the nearest neighborhood}

\label{sec-local}

In this section we focus on how local magnetic moments depend on the
chemical composition of the nearest neighborhood.  All values
presented here were obtained for non-relaxed structures via the
\kkr\ code (as in the previous sections). 
                
\begin{figure}
\includegraphics [width=0.9\linewidth]{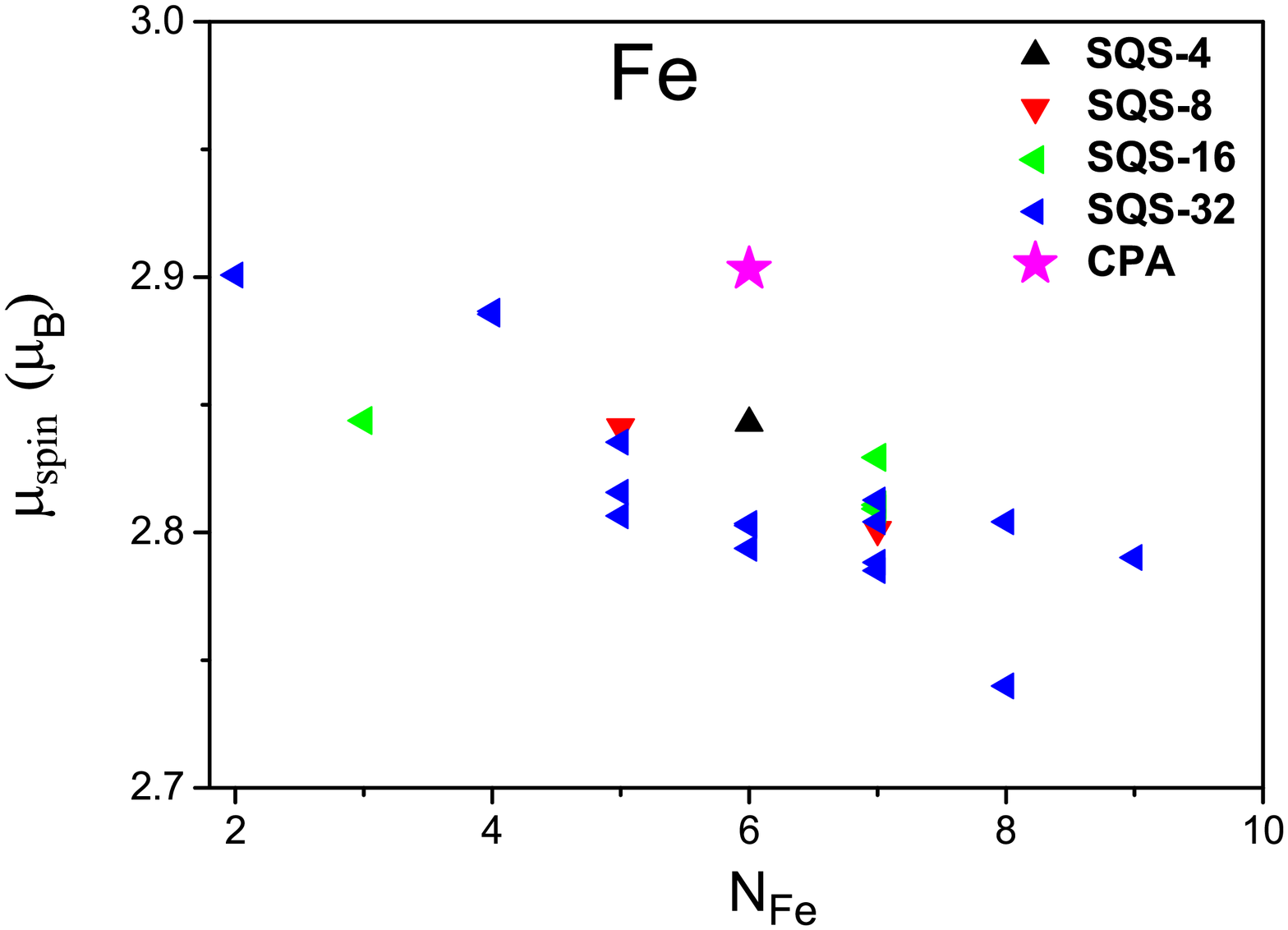} 
 \includegraphics [width=0.9\linewidth]{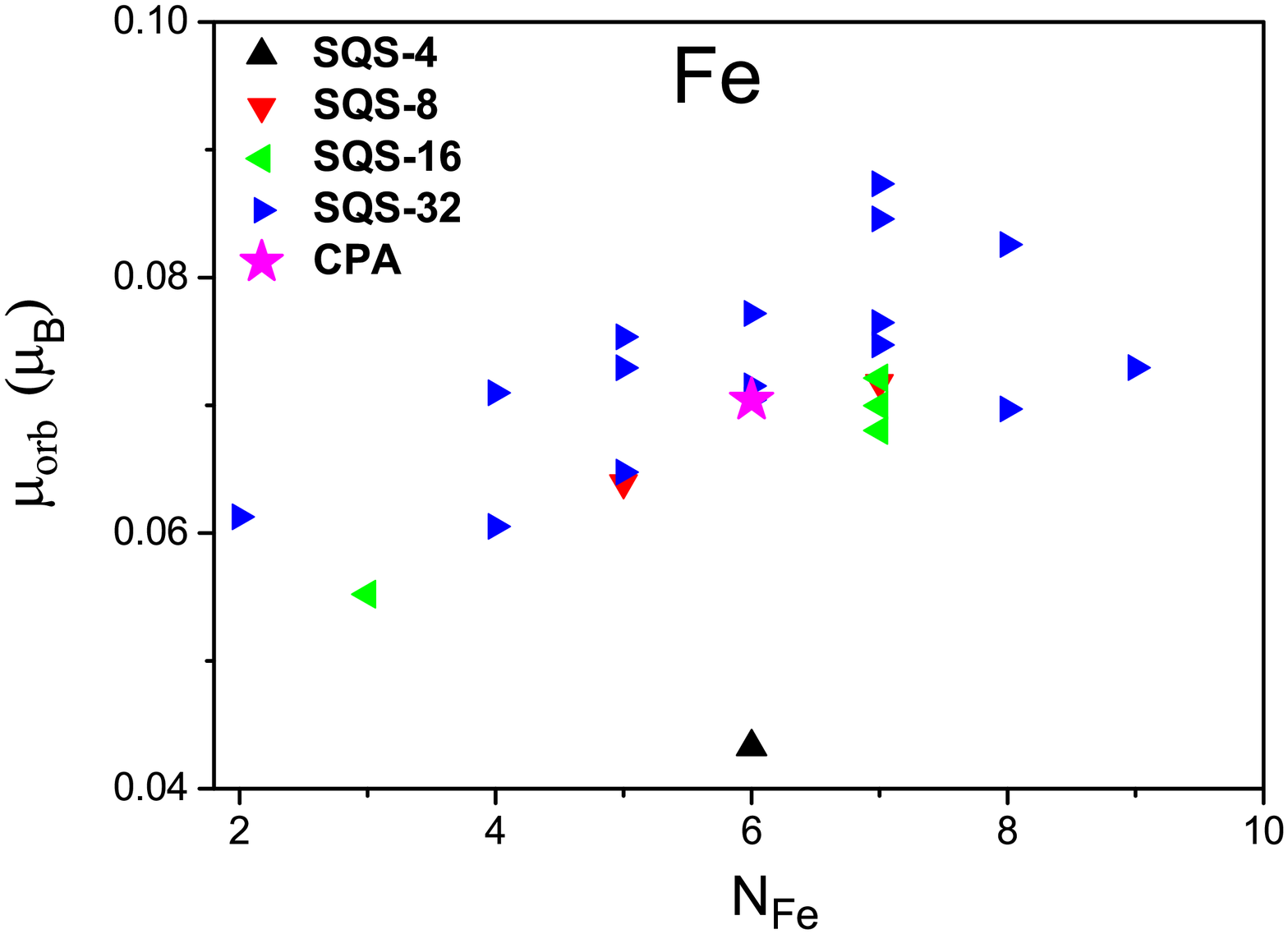}
\caption{Spin and orbital magnetic moments for Fe sites in various
  SQS's shown as a function of the number of Fe atoms in the first
  coordination sphere. The CPA results are shown for comparison.  The
  data were obtained by the \kkr\ code.}
 \label{Fe-coordination-fig}
\end{figure}

Figure~\ref{Fe-coordination-fig} shows the local spin and orbital
magnetic moments, \ms\ and \mo, respectively, for Fe sites in each of
the SQS's as a function of the number of Fe atoms in the first
coordination sphere, \nfe. The magnetization is always parallel to the
[001] direction of the parental fcc lattice.

One can see from Fig.~\ref{Fe-coordination-fig} that the values of
\ms\ for the Fe sites are all very similar --- they do not differ from
each other by more than 5~\%.  There is a much larger spread for the
local \mo\ values (around 30~\%).  Generally, \ms\ for Fe atoms
decreases with increasing \nfe. This is plausible, because increasing
\nfe\ means strengthening the hybridization between Fe states, which
suppresses the magnetic moment.  An opposite trend is observed for
\mo. This is a bit surprising because, usually, \mo\ exhibits the same
trend as \ms\ for 3$d$ atoms \cite{SKE+04,BSM+12}.  The explanation
rests on the large SOC at Pt atoms in comparison with the SOC at Fe
atoms: \mo\ at Fe atoms is suppressed by the off-site SOC at Pt atoms.
This mechanism was discussed in detail for the CoPt system
\cite{Sipr+08} and apparently is active here as well.  Indeed, the
trends of \ms\ and \mo\ for Fe atoms in FePt are quite similar to the
trends of \ms\ and \mo\ for Co atoms in CoPt systems (compare our
Fig.~\ref{Fe-coordination-fig} with figures~1 and~4 of \v{S}ipr
\ea\ \cite{Sipr+08}).

\begin{figure}
\includegraphics [width=0.9\linewidth]{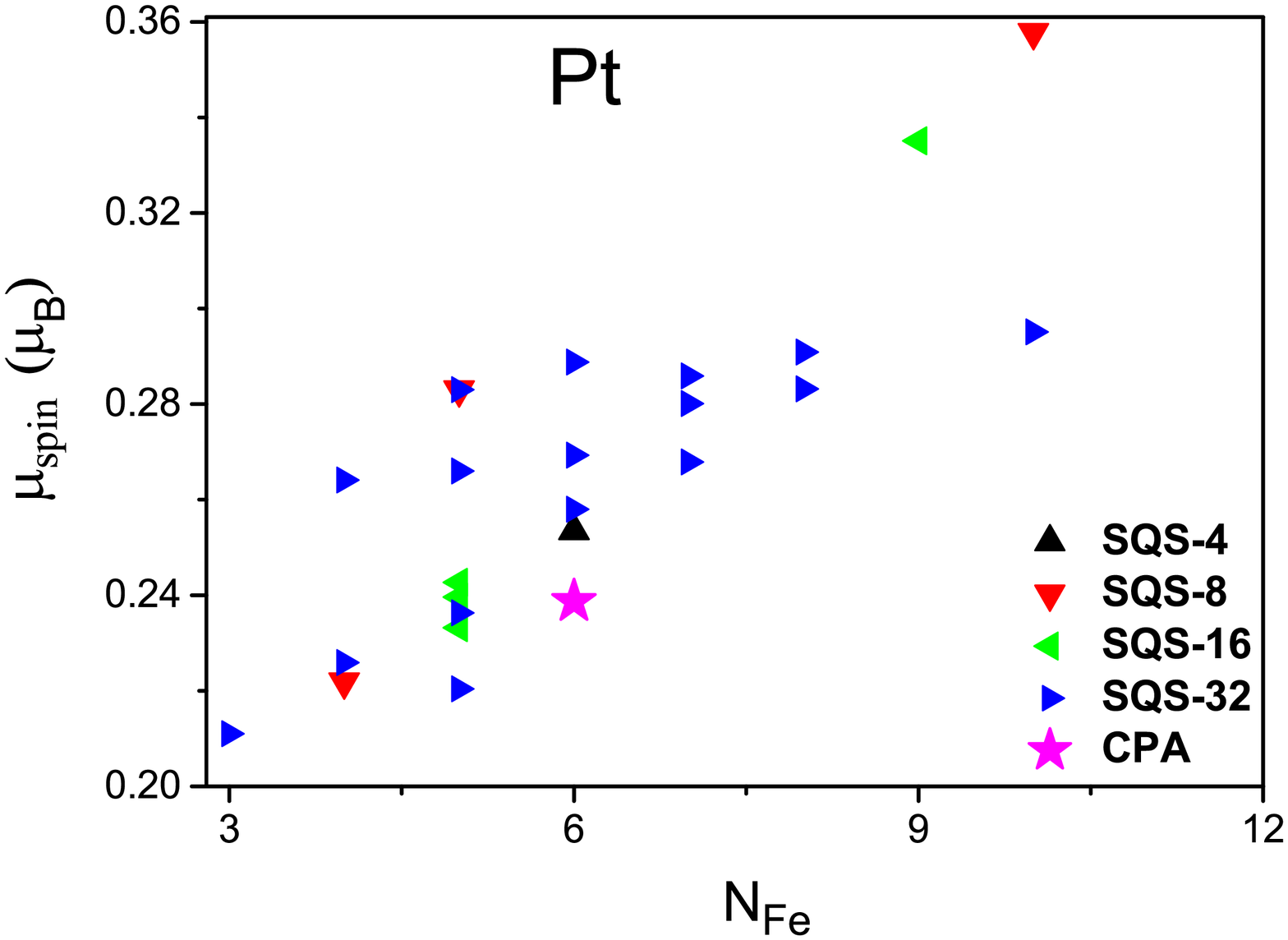}
\includegraphics [width=0.9\linewidth]{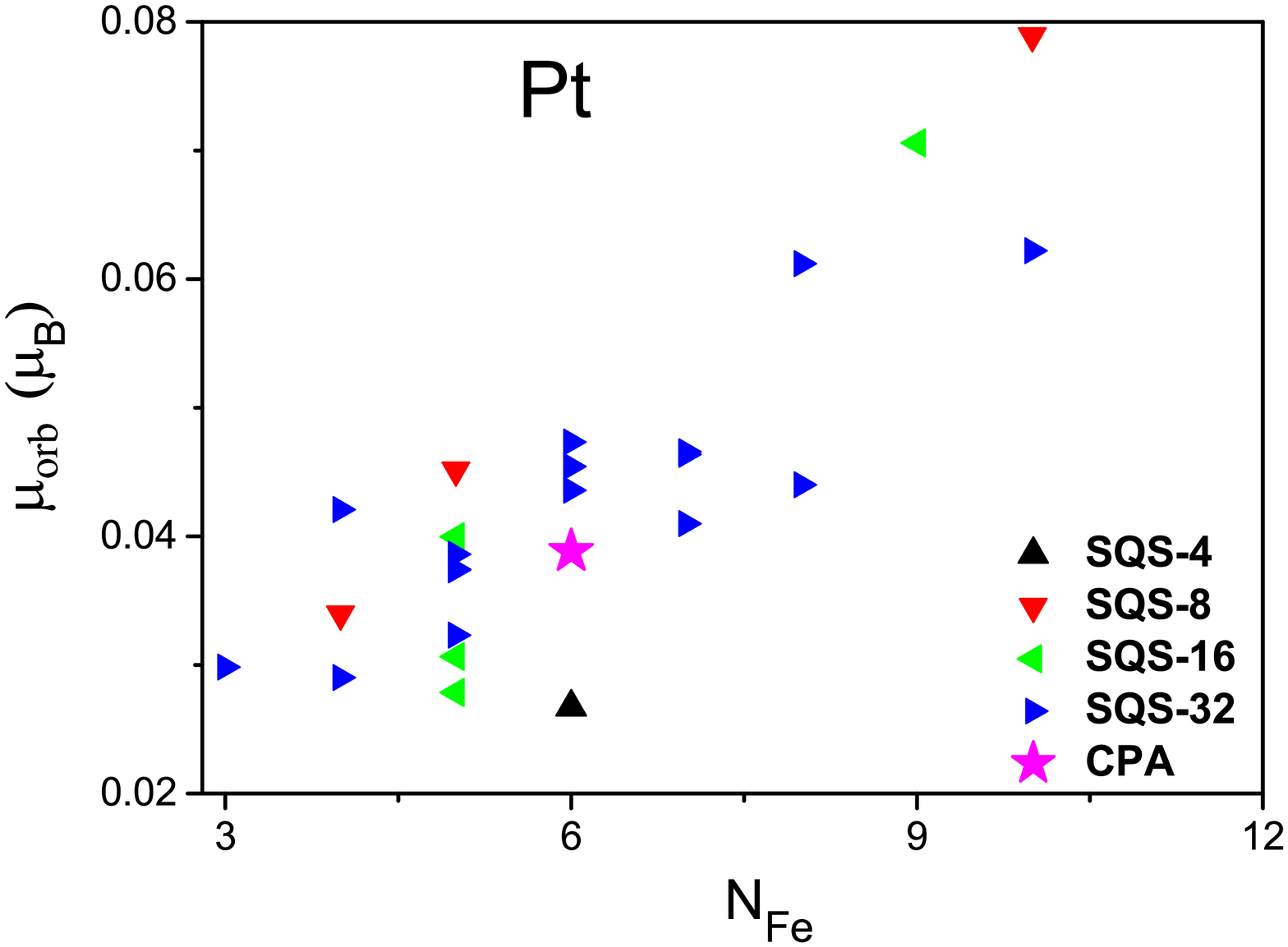}
\caption{As Fig.~\protect\ref{Fe-coordination-fig} but for Pt
  sites. }
\label{Pt-coordination-fig}
\end{figure}

The situation is different concerning the moments for the Pt sites.
Corresponding local moments \ms\ and \mo\ are shown as a function of
\nfe\ in Fig.~\ref{Pt-coordination-fig}.  One can see that if the
number of Fe atoms near a Pt atom increases, \ms\ and \mo\ for this Pt
atom increase as well.  This is consistent with the fact that the
magnetic moments of the Pt atoms are induced by the magnetic moments
of the neighboring Fe atoms.  If the number of neighboring Fe atoms
increases, so does the induced magnetic moment at the Pt site.  This
is true both for 
\ms\ and \mo; the disturbance by the off-site SOC at neighboring Fe
atoms is not significant because the SOC-strength parameter for Pt
atoms (712~meV as obtained by the expression of Davenport
\ea\ \cite{DWW+88}) is much larger than the SOC-strength parameter for
Fe atoms (65~meV).

\begin{figure}
\includegraphics [width=0.9\linewidth]{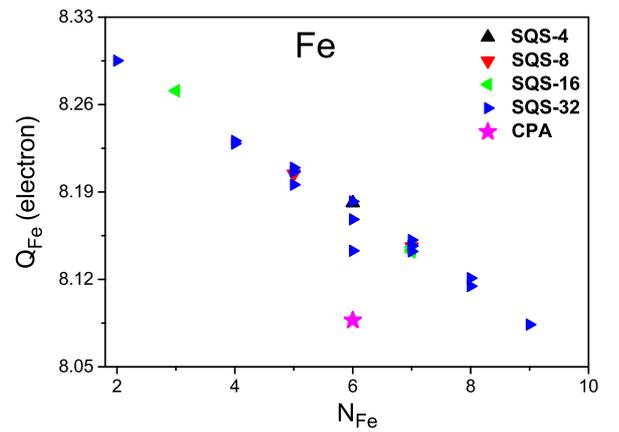} 
\caption{Electronic charge for Fe sites in various SQS's shown as a
  function of the number of Fe atoms in their first coordination
  spheres.  The CPA result is shown for comparison.  The data were
  obtained by the \kkr\ code.}
 \label{Fe-charge}
\end{figure}

The CPA leads to different \ms\ than what is obtained for the
supercells (Fig.~\ref{Fe-coordination-fig}).  To find why it is so, we
look at the dependence of the electronic charge at the Fe site
($Q_{\text{Fe}}$) on \nfe, again for various SQS's.  The corresponding
graph is given in Fig.~\ref{Fe-charge}.  Here, a convincing
quasi-linear relation between $Q_{\text{Fe}}$ and \nfe\ can be seen.
One sees, at the same time, that the CPA result clearly does not fit
into the trend set by the supercells.  An analogous plot could be
drawn also for electronic charge at the Pt sites (just with an
opposite trend).

An interesting aspect following from Fig.~\ref{Fe-charge} is that the
Fe atoms actually gain electrons when they are alloyed with Pt.  Of
course, this depends on the way the atomic regions are defined; in our
case, we use Voronoi polyhedra, meaning that Fe and Pt atoms occupy
identical volumes.  The charge flow picture might look differently for
different definitions of the atomic cells.

\begin{figure}
\includegraphics [width=0.9\linewidth]{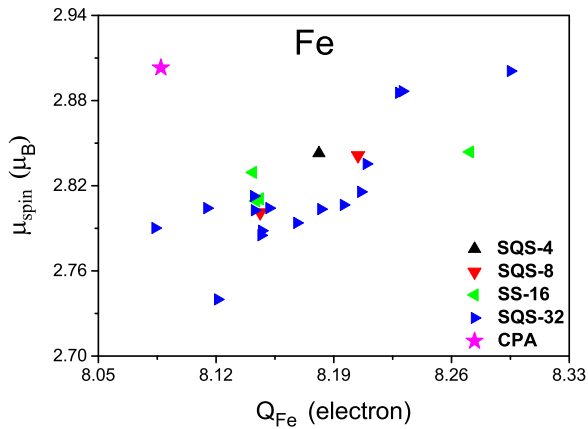}
\caption{Spin magnetic moment for Fe sites in various SQS's shown as
  a function of the charge.  The CPA result is shown for comparison.
  The data were obtained by the \kkr\ code. }
 \label{mag-charge}
\end{figure}

For a complete view we explore also the dependence of the magnetic
moments at the Fe sites on the charge $Q_{\text{Fe}}$
(Fig.~\ref{mag-charge}).  Similarly as for Fig.~\ref{Fe-charge}, the
data points for the supercells exhibit a common trend and the CPA
stands clearly out of it.


\subsection{Influence of the Madelung potential} 

\label{sec-madel} 

The CPA leads to significantly different magnetic moment and charge
than what would correspond to a Fe atom in a supercell with
$N_{\text{Fe}}=6$ (figures~\ref{Fe-coordination-fig} and
\ref{Fe-charge}).  It thus appears that there is a difference between
the way magnetism in disordered FePt alloy is described via the CPA
and via the supercell approach and that this difference is linked to
the electronic charge.  A possible reason for this difference is the
single-site nature of the CPA.  In particular, the standard CPA cannot
account for the Madelung contribution to the potential.  To get more
detailed insight, we investigate to what extent including or ignoring
the Madelung potential affects the electronic structure and magnetism
of the SQS's.
                
\begin{figure}
\includegraphics [width=0.9\linewidth]{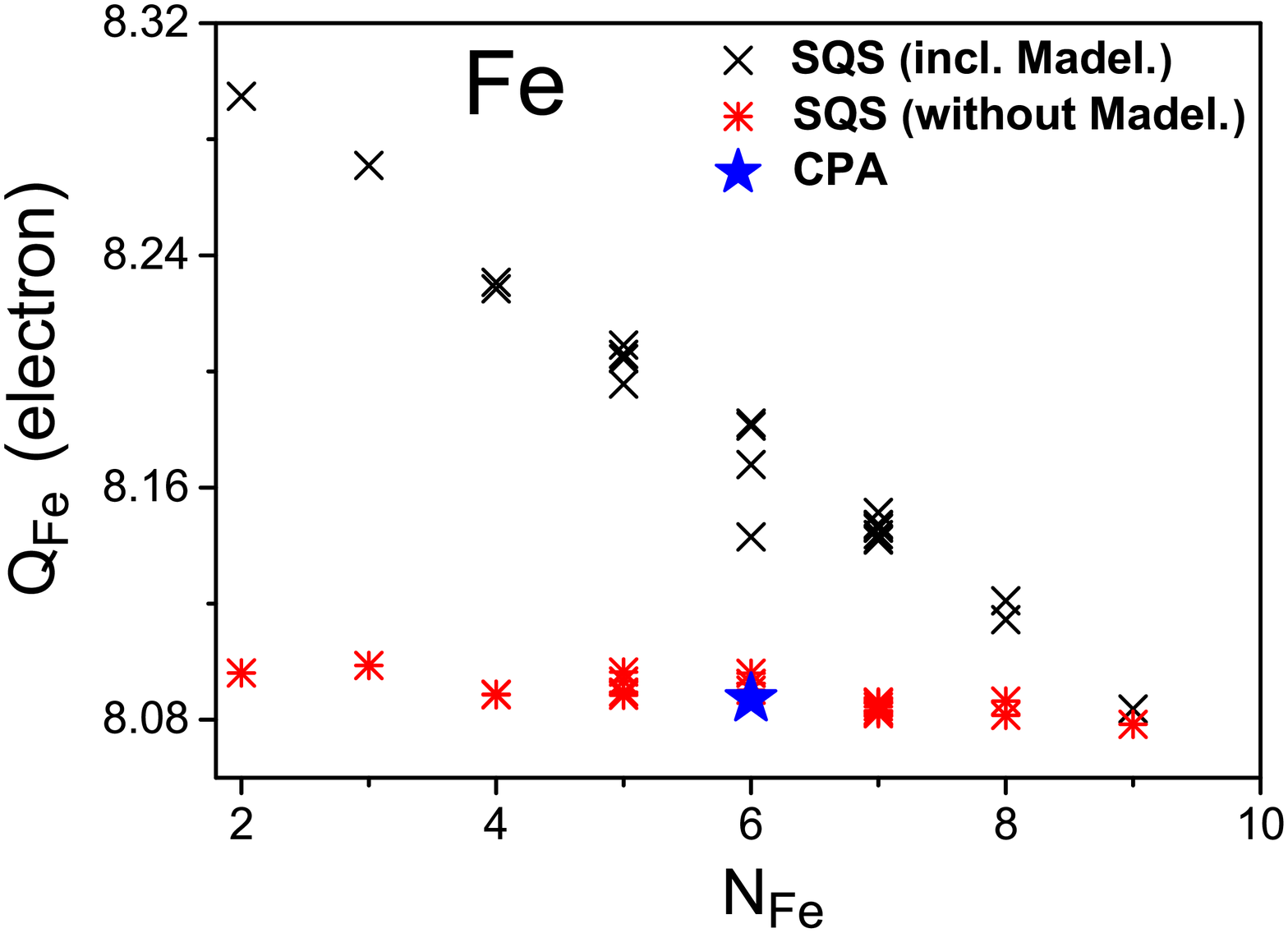}
\includegraphics [width=0.9\linewidth]{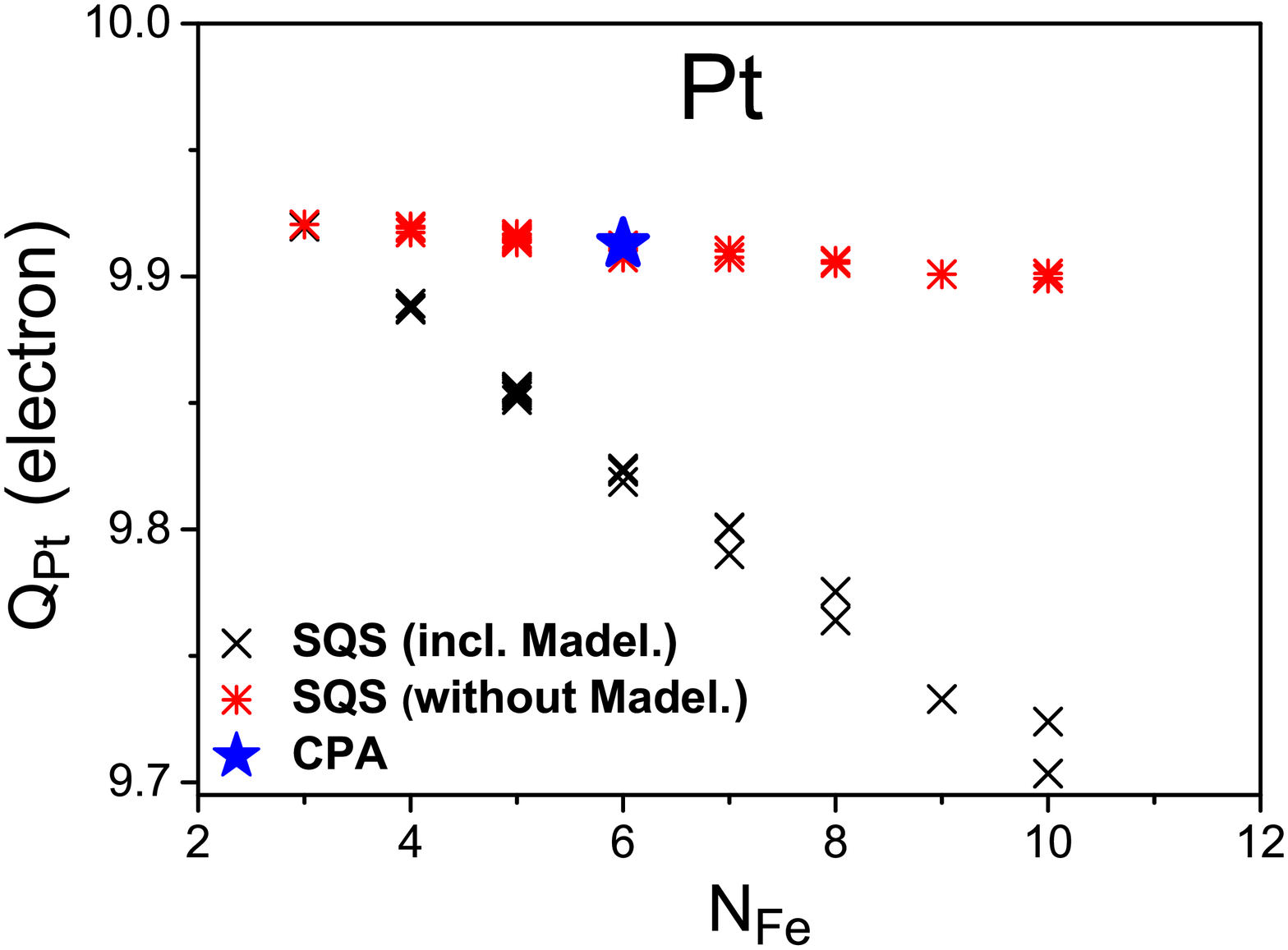}
\caption{Electronic charge at the Fe sites (upper panel) and at the Pt
  sites (lower panel) as function of \nfe\ for SQS-4, 8, 16 and 32
  obtained with the Madelung potential either included or ignored.
  The CPA result is shown as well.  The calculations were done using
  the \kkr\ code.}
 \label{chg-madel-fig}
\end{figure}

The influence of the Madelung potential on the charges at the Fe sites is
shown in Fig.~\ref{chg-madel-fig}.  One can see immediately that
neglecting the Madelung potential practically suppresses the linear
dependence of the charge on the coordination number.  The CPA result
corresponds to the case when the Madelung potential is neglected.  For
the Pt sites the situation is similar as for the Fe sites.
                
\begin{figure}
\includegraphics [width=0.9\linewidth]{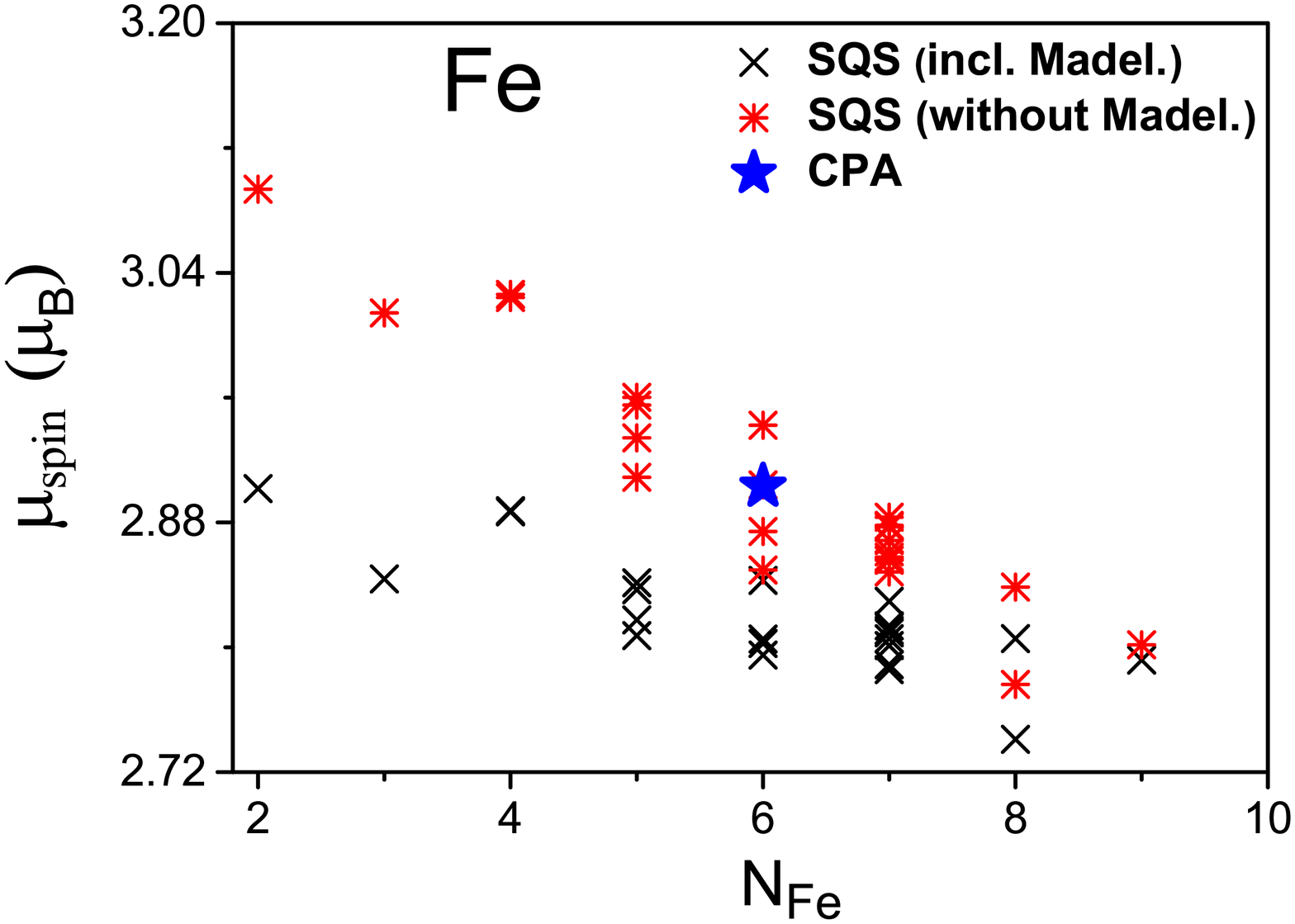}
\includegraphics [width=0.9\linewidth]{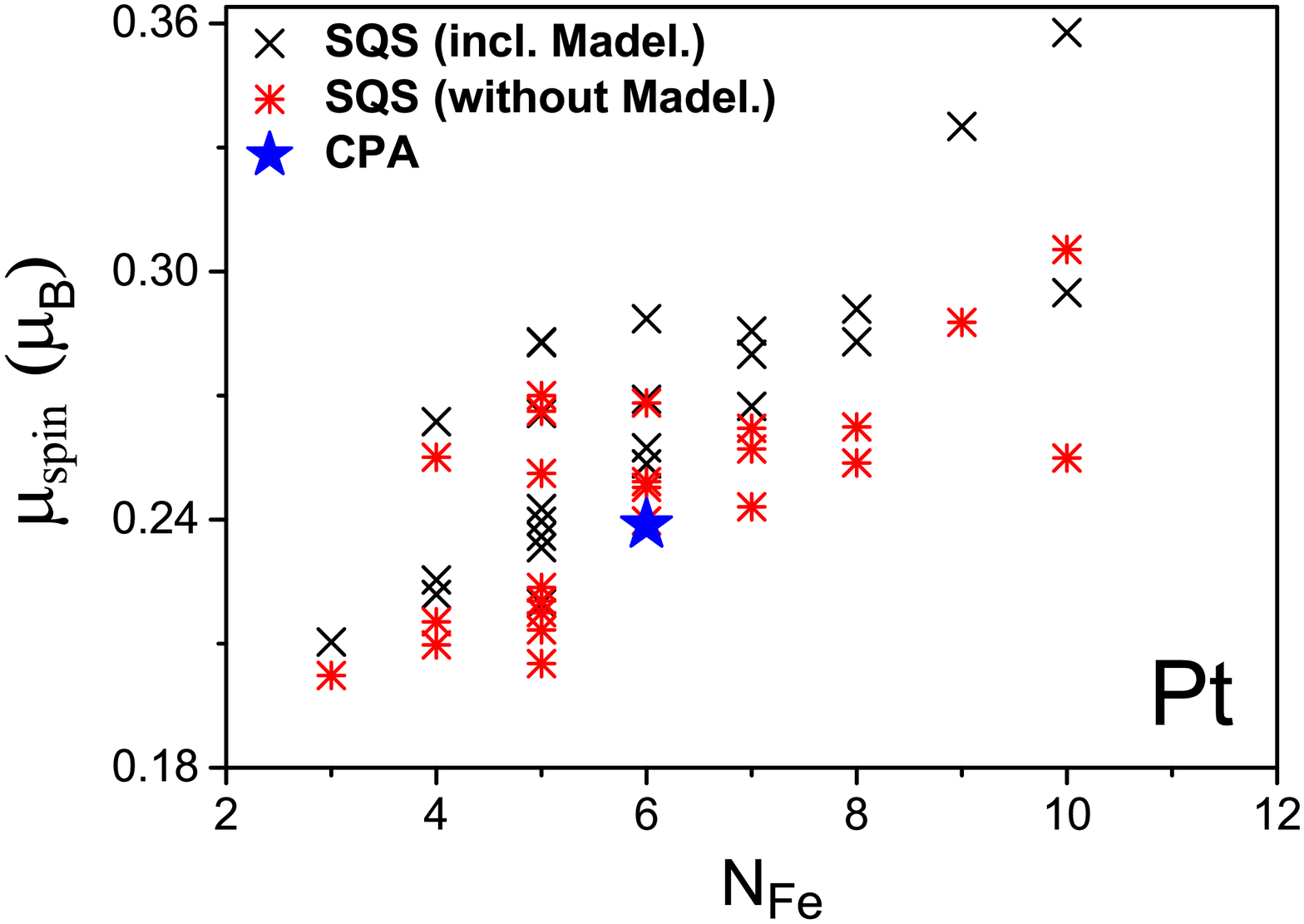}
\caption{As figure \protect\ref{chg-madel-fig} but for spin magnetic
  moments.}
 \label{spn-madel-fig}
\end{figure}

An analogous comparison for the spin magnetic moments is presented in
Fig.~\ref{spn-madel-fig}.  Even though the difference between the
situation with the Madelung potential and without it is not so
striking as in Fig.~\ref{chg-madel-fig}, again we see that the data
split into two groups, with different slopes.  This is true especially
for the Fe sites: Neglecting the Madelung potential increases \ms\ by
0.05--0.10~\mB, which is about the same as the difference between
\ms\ for Fe obtained by averaging over all SQS's and by the CPA
(Table~\ref{average-moments}).  Interestingly, the spin magnetic
moments at the Pt sites are less affected by the presence or absence
of the Madelung potential (see Fig.~\ref{Pt-coordination-fig} and also
Table~\ref{madel-pot}).  This is probably linked to the fact that the
magnetic moments at the Pt sites are induced by the moments at the Fe
sites, so the effect of the Madelung potential is felt not directly at
the Pt sites themselves but only indirectly, through the change of the
moments at the Fe sites.

\begin{table} 
\caption{ Comparison of charge $Q$ (in electrons) and \ms\ (in
  \mbox{$\mu_{B}$}) obtained by averaging over all sites in the SQS-4,
  SQS-8, SQS-16 and SQS-32 supercells (separately for Fe and Pt atoms,
  either with or without the Madelung potential) with values obtained
  for disordered FePt via the CPA. The upper part of the table
  contains the full potential results and lower part contains the ASA
  results. \label{madel-pot}}
\begin{ruledtabular}
\begin{tabular}{ldddd}
& \multicolumn{2}{c}{Fe} & \multicolumn{2}{c}{Pt} \\ 
    &
  \multicolumn{1}{c}{$Q_{\mathrm{Fe}}$}             &
  \multicolumn{1}{c}{\mbox{$\mu_{\mathrm{spin}}$}}   &
  \multicolumn{1}{c}{$Q_{\mathrm{Pt}}$}             &
  \multicolumn{1}{c}{\mbox{$\mu_{\mathrm{spin}}$}}   \\
\hline 
full potential:      &    &   &   &   \\ 
$\left<\text{SQS}\right>$ incl.\ Madelung
               &  8.174  &  2.821  &  9.825  &  0.266  \\ 
$\left<\text{SQS}\right>$ without Madel.\
               &  8.088  &  2.907  &  9.912  &  0.245  \\
CPA            &  8.087  &  2.903  &  9.913  &  0.239  \\[1.2ex]  
ASA:       &    &   &   &   \\ 
$\left<\text{SQS}\right>$ incl.\ Madelung
               &  8.150  &  2.864  &  9.850  &  0.247  \\ 
$\left<\text{SQS}\right>$ without Madel.\
               &  8.083  &  2.929  &  9.917  &  0.240  \\
 CPA           &  8.083  &  2.928  &  9.917  &  0.232  \\ 
\end{tabular}
\end{ruledtabular}
\end{table}

Table~\ref{madel-pot} summarizes how the Madelung potential affects
the charge and the spin magnetic moment averaged over all sites of given
atomic type.  As there can be some ambiguity how the suppression of
the Madelung potential should be technically performed in the full
potential case, we include in Table~\ref{madel-pot} also the results
for the ASA.  The outcome is similar in both cases: If the Madelung
potential is suppressed, averaging over the supercells yields
practically the same charges and magnetic moments as the CPA.

Several modifications of the CPA method were proposed to account for
the effect of the Madelung potential \cite{Johnson+93, Magri+90,
  Korzhavyi+95, Saha+96, Ujfalussy+00, Bruno+02, Abrikosov+98,
  Ruban+02}.  A survey of these approaches would be beyond our scope
but to make yet another assessment of the influence of the Madelung
potential, we employ the screened impurity model for the Madelung
contribution within the CPA (SIM-CPA) \cite{Ruban+02}.  This model
assumes that the Madelung potential can be modeled as the potential
due to a screening charge spherically distributed at the
nearest-neighbor distance.  Using this approach, we obtain
2.853~\mB\ for \ms\ at Fe atom and 0.244~\mB\ for \ms\ at Pt atom.
Comparing these values with Table~\ref{average-moments}, we see that
the SIM-CPA method pushes the standard CPA results in the desired
direction.  By optimizing model parameters of the SIM-CPA method
\cite{Ruban+03}, these values could be brought even closer to the
values obtained by averaging over the SQS's.  However, this would not
bring any new physical insight.  We conclude that neglecting the
Madelung potential by the CPA leads to small but distinct changes in
magnetic moments in FePt alloy.


\subsection{Dependence of local magnetic moments on bond lengths} 

\label{sec-lengths} 

By optimizing the positions of the atoms in the SQS, a variety of bond
lengths is obtained.  Here we study how variations in bond lengths
affect magnetic moments.  Respective calculations were performed using 
the \wien\ code.

Generally, two types of structural relaxations could be made for an
SQS: relaxation of internal degrees of freedom and relaxation of
external degrees of freedom.  Relaxation of internal degrees of
freedom means that atoms are allowed to move in the direction of a
force whereas the lattice vectors are kept unchanged.  Relaxation of
external degrees of freedom means that the lengths of the lattice
vectors and the angles between them are optimized.  External degrees
of freedom reflect whole manifold of possible configurations, it is
thus reasonable to keep them fixed when using the supercell to model
an alloy.  To study local environment effects, we relax only the
atomic positions.

\begin{figure}
\includegraphics [width=0.51\linewidth]{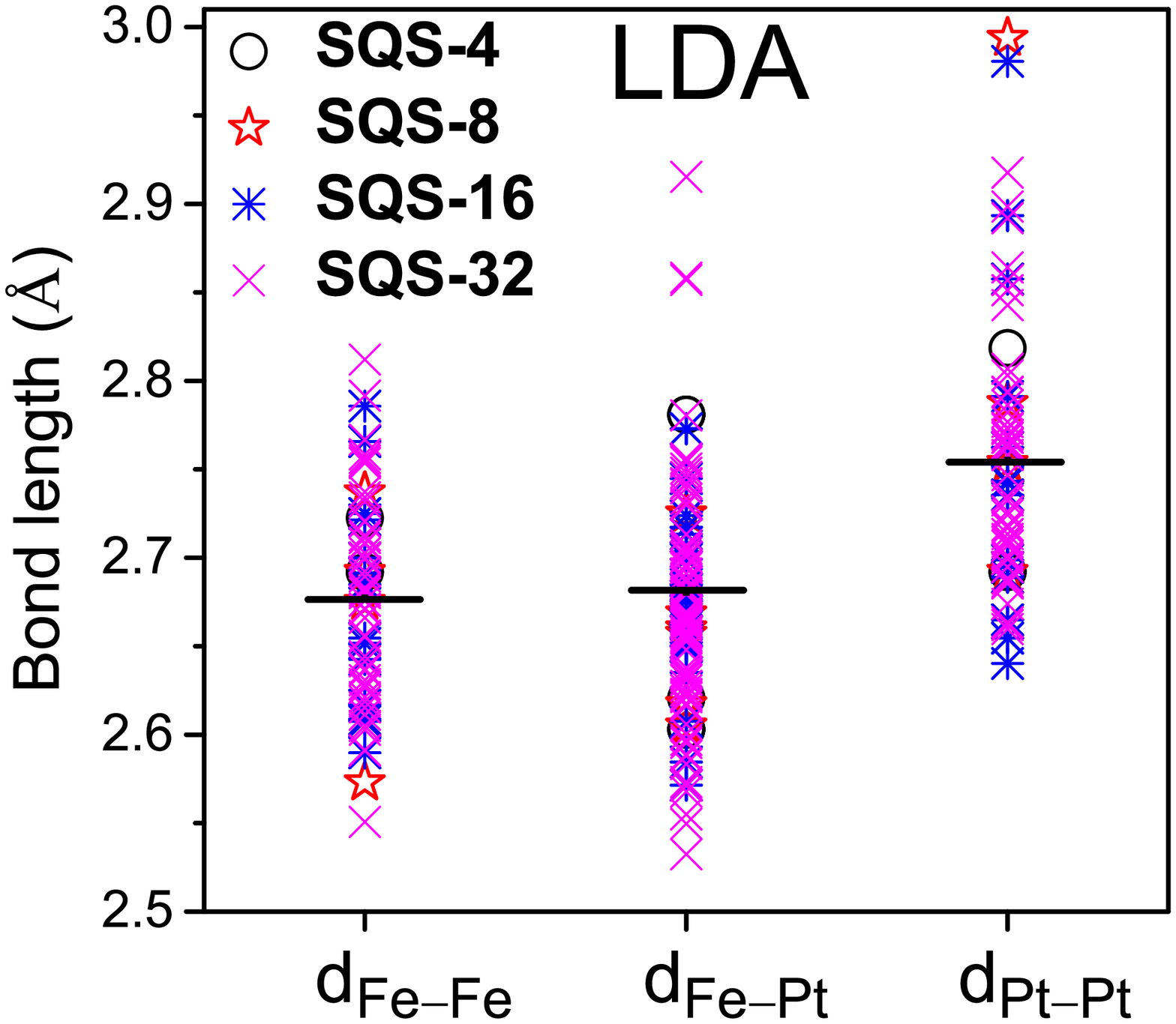} 
\includegraphics [width=0.46\linewidth]{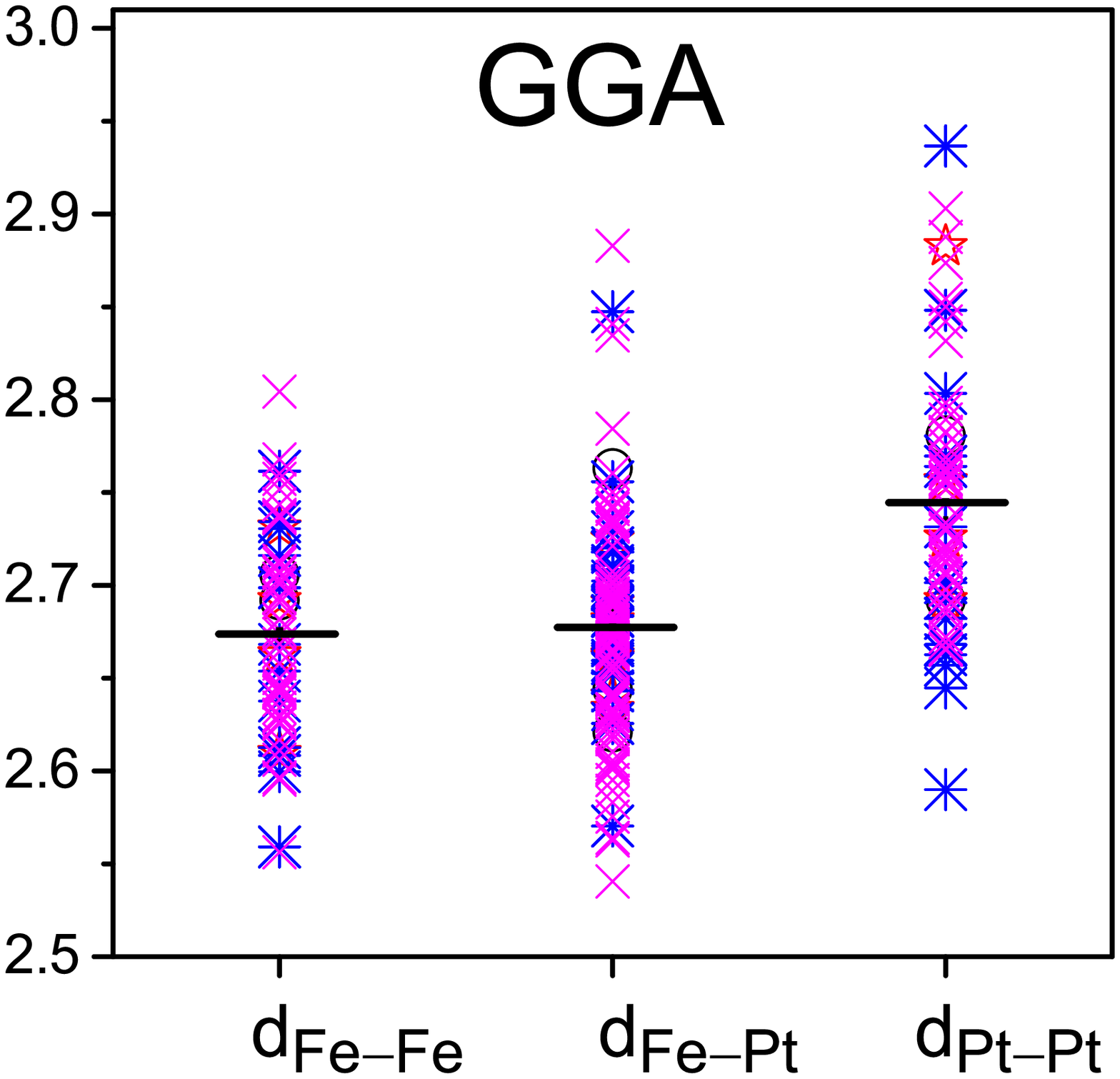} 
\caption{Optimized inter-atomic distances for the first coordination
  shell in SQS-4, 8, 16, and 32.  Average values are shown by
  horizontal lines.  Results were obtained using the \wien\ code, with
  the exchange-correlation potential parametrized within the LDA (left
  panel) and within the GGA (right panel).}
\label{bond-length}
\end{figure}

Bond-lengths $d_{\text{Fe--Fe}}$, $d_{\text{Fe--Pt}}$ and
$d_{\text{Pt--Pt}}$ resulting from the geometry optimization of the
internal degrees of freedom are shown in Fig.~\ref{bond-length} (both
for the LDA and for the GGA).  For a non-relaxed structure, all the
lengths are 2.69~\AA.  One can see that the $d_{\text{Fe--Fe}}$ and
$d_{\text{Fe--Pt}}$ distances are on the average close to 2.69~\AA,
with $d_{\text{Fe--Fe}}$ a bit smaller than $d_{\text{Fe--Pt}}$.  The
$d_{\text{Pt--Pt}}$ distances are on the average larger than
$d_{\text{Fe--Fe}}$ or $d_{\text{Fe--Pt}}$ distances.  This is
consistent with the fact that the inter-atomic distances in elemental
Pt (2.77~\AA) are significantly larger than the inter-atomic distances
in elemental Fe (2.48~\AA).  The overall pictures provided by the LDA
and by the GGA are similar.  In the following only results for the LDA
will be presented.

\begin{figure}
\includegraphics [width=0.9\linewidth]{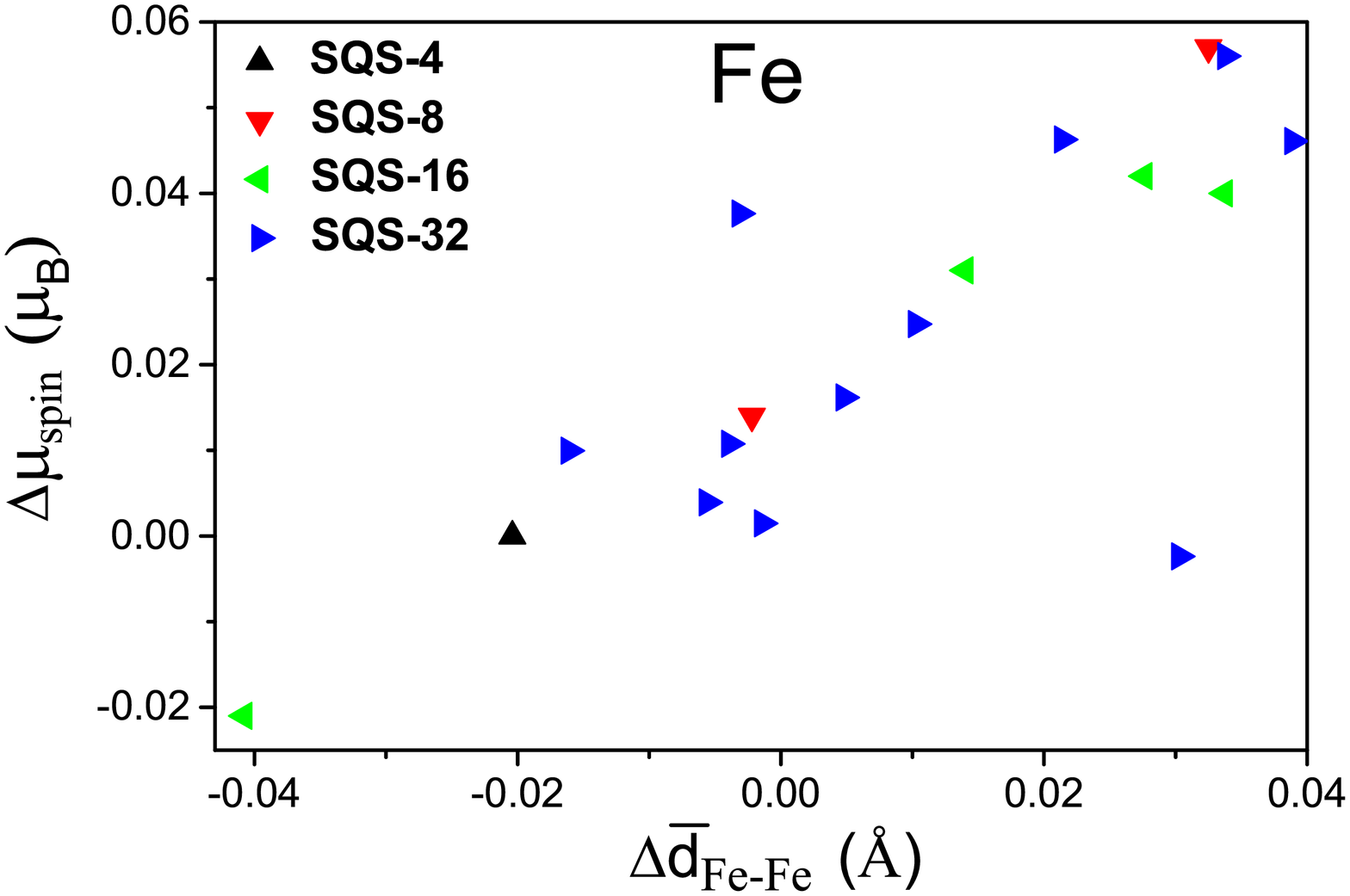}
\includegraphics [width=0.9\linewidth]{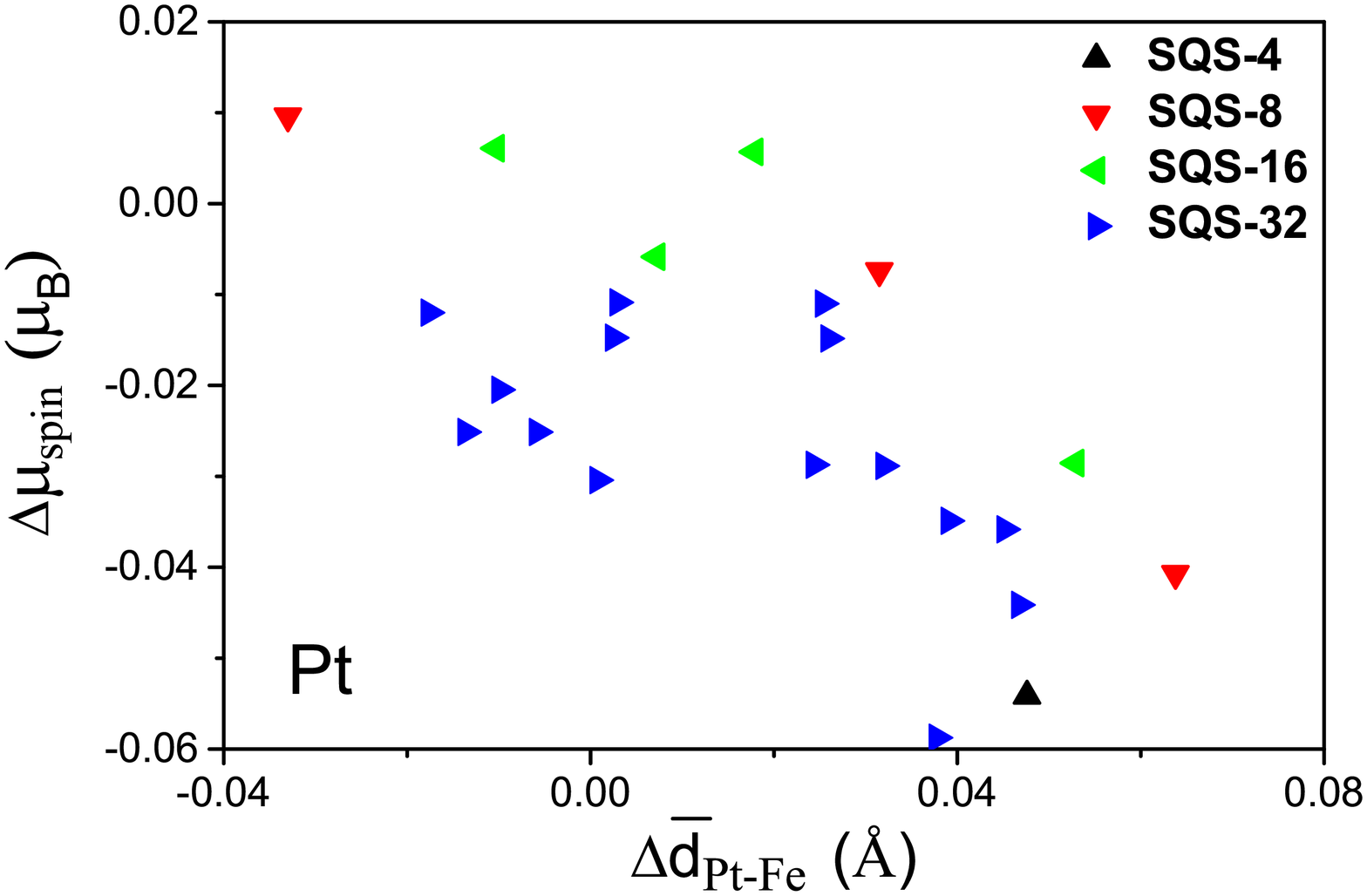}
\caption{Change of the spin magnetic moment $\Delta\mu_{spin}$ for a
  Fe atom (upper graph) and for a Pt atom (lower graph) plotted as a
  function of change of the average bond length $\Delta
  \bar{d}_{\text{Fe-Fe}}$ or $\Delta \bar{d}_{\text{Pt-Fe}}$. The
  calculations were done by the \wien\ code.}
 \label{average-bond-length}
\end{figure}

Changes in the inter-atomic distances cause corresponding changes in
the magnetic moments.  It is instructive to inspect how the change in
the local spin moment \ms\ is related to the change in the average
distance of the nearest Fe neighbors from the respective site ($\Delta
\bar{d}_{\text{X-Fe}}$).  This is shown in
Fig.~\ref{average-bond-length}.  One can see that if the Fe neighbors
around a Fe site are pushed away (i.e., $\Delta
\bar{d}_{\text{Fe-Fe}}$ increases), \ms\ for that site increases.
This is plausible, because increasing $\Delta \bar{d}_{\text{Fe-Fe}}$
means that the hybridization between states related to Fe atoms
decreases leading to an enhancement of the magnetic moment.  On the
other hand, if the Fe neighbors around a Pt site are pushed away
(i.e., $\Delta \bar{d}_{\text{Pt-Fe}}$ increases), \ms\ for this Pt
atom decreases.  This reflects the fact that the magnetic moments at
Pt atoms are induced by neighboring Fe atoms; the effectiveness of
this mechanism obviously decreases with increasing Pt--Fe distance.


\section{Discussion}

\label{Discussion}

Our goal was to compare the CPA and supercell description of
electronic structure and magnetism of disordered FePt and to search for
effects of the local environment.  We found that the DOS averaged over
all sites in supercells quickly approaches the DOS provided by the
CPA.  This means that although there may be large variations between
the DOS for individual sites (as highlighted, e.g., by Lu
\ea\ \cite{Lu+91}), these variations are smeared out if an average is
taken over even a small number of modestly large supercells.  This was
observed earlier for CuPd, CuAu and CuZn alloys \cite{AJ+98} and for
NiPt alloy \cite{RSK+02}.  The good agreement between the energy
dependence of the DOS obtained by averaging over the supercell and by
the CPA suggests that the CPA is very efficient for calculating
spectroscopic properties \cite{GMU+12,SVS+16}.

We found that the substitutional disorder in FePt can be described by
relatively small supercells.  The SQS-16 structure provides already a
good description. Similar conclusions were reached earlier for a NiPt
alloy by Ruban \ea\ \cite{RSK+02} and by Shang \ea\ \cite{Shang+11}.

Concerning the magnetism we found that the CPA leads to magnetic
moments that differ from the limit approached by involving supercells
of ever increasing sizes.  The difference between the CPA and the
supercell limit is about 0.08~\mB\ for \ms\ at the Fe atoms and about
-0.03~\mB\ for \ms\ at the Pt atoms.  The reason for this difference
is the neglect of the Madelung contribution to the potential in the
standard CPA formalism.

Our study of the influence of the chemical composition of the nearest
neighborhood on local magnetic moments can be seen as an extension of
earlier studies of the influence of the coordination number on the
local magnetic moments for 3$d$ transition metal clusters and surfaces
\cite{SKE+04, MLZ+06,BSM+12}. Similarly as for transition metals
clusters and surfaces, a link between local magnetic moments for Fe
atoms and the number of their nearest Fe neighbors can be established
(Fig.~\ref{Fe-coordination-fig}) --- albeit with a large spread around
the purported dependence. The fact that the link between \ms\ and
\nfe\ is less evident for FePt than for Fe clusters indicates that the
decrease of hybridization caused by replacing a neighboring Fe atom by
a Pt atom is smaller than the decrease caused by removing a
neighboring atom altogether.  This is plausible.

The link between magnetic moments for Pt atoms and the number of their
Fe neighbors \nfe\ appears to be quite strong (figure
\ref{Pt-coordination-fig}).  This link stresses that the magnetism of
the Pt atoms is induced by their Fe neighbors.  Yet another
demonstration of this fact comes from the dependence of the magnetic
moment for Pt on the $d_{\text{Fe--Pt}}$ distance.  The results shown
in Fig.~\ref{average-bond-length} are in accordance with an earlier
study where the magnetic moments of Pt atoms were treated as
originating solely from nearest magnetic 3$d$ atoms~\cite{PMS+10}.

We found that the electronic charge at Fe sites decreases with the
number of Fe atoms in the first coordination shell.  The opposite is
true for Pt sites.  This can be reformulated that the excess charge at
Fe or Pt sites increases linearly with the number of unlike atoms in
their nearest neighborhoods.  Similar observations were made earlier
for AgPd and AgAu alloys \cite{Lu+91} and for CuZn alloys
\cite{JP+93}.

Concerning the fluctuations of the bond lengths, we note that the
spread of the bond lengths due to structural relaxation shown in
figure \ref{bond-length} is approximately the same as found by Ruban
\ea\ \cite{Ruban+03} and by Shang \ea\ \cite{Shang+11} for NiPt.


\section{Conclusions} \label{Conclusions}

Our calculations show that by increasing the number of atoms $N$ in
special quasirandom structures SQS-$N$, the results for the magnetic
moments and for the DOS approach the results obtained via the CPA.
However, a small but distinct residual difference remains between
magnetic moments obtained by both approaches.  This is due to the
neglect of the Madelung potential in the standard CPA.

The local magnetic moments associated with Fe atoms are more robust
with respect to variations of the local environment than the magnetic
moments associated with the Pt atoms.  This reflects the fact that
magnetism at Pt sites is induced by neighboring Fe atoms and that
electronic states derived from the Pt atoms are more delocalized than
states derived from the Fe atoms.

If structural relaxation is performed, the distances between the Pt
atoms $d_{\text{Pt--Pt}}$ are larger on the average than distances
between the Fe atoms $d_{\text{Fe--Fe}}$ or distances between the Fe
and Pt atoms $d_{\text{Fe--Pt}}$.  The magnetic moments at the Fe
sites increase if the average $d_{\text{Fe--Fe}}$ distance increases.
The magnetic moments at the Pt sites decrease if the average
$d_{\text{Fe--Pt}}$ distance increases, in accordance with intuition.

If the disorder in FePt alloy is simulated by supercells constructed
along the SQS prescription, the calculated integral properties
converge with the size of the SQS quite quickly: accuracy sufficient
for most needs is achieved already for the SQS-16 supercell.


\begin{acknowledgments}
We would like to acknowledge CENTEM project (CZ.1.05/2.1.00/03.0088),
CENTEM PLUS (LO1402) and COST CZ LD15147 of The Ministry of Education,
Youth and Sports (Czech Republic). Computational time has been provided with the
MetaCentrum (LM205) and CERIT-SC (CZ.1.05/3.2.00/08.0144)
infrastructures.  In addition we would like to thank for travel
support from EU-COST action MP1306 (EUSpec).
\end{acknowledgments}




\bibliography{biblo-sqs}


\end{document}